\documentclass[12pt]{article}

\input epsf
\usepackage[dvips]{graphicx}
\usepackage{amsfonts,amsmath,amssymb}

\newcommand{\pa}{\partial}
\newcommand{\nn}{\nonumber}

\newcommand{\cA}{{\cal A}}

\def\href#1#2{#2}

\begin{document}

\begin{titlepage}

\begin{center}

\hfill 
\vskip 4cm

\textit{\Large Excursions through KK modes}\\[14mm] 

{\it Kazuyuki Furuuchi}\\[8mm]

{\it Manipal Centre for Natural Sciences, Manipal University}\\
{\it Manipal, Karnataka 576104, India}

\vskip8mm 
\end{center}
\begin{abstract}
In this article we study
Kaluza-Klein (KK) dimensional reduction of 
massive Abelian gauge theories with 
charged matter fields on a circle.
Since local gauge transformations 
change position dependence of the charged fields,
the decomposition of the charged matter fields
into KK modes is gauge dependent.
While whole KK mass spectrum
is independent of the gauge choice,
the mode number depends on the gauge.
The masses of the KK modes also depend on the 
field value of the zero-mode of 
the extra dimensional component of the gauge field.
In particular,
one of the KK modes in the KK tower of each massless 5D charged field
becomes massless at particular values of 
the extra-dimensional component 
of the gauge field.
When the extra-dimensional component of the gauge field
is identified with the inflaton,
this structure leads to recursive
cosmological particle productions.
\end{abstract}

\end{titlepage}

\section{Introduction}\label{sec:Intro}

Local gauge transformations
change the position dependence of charged matter fields.
As a result, when
compactified on a circle,
the decompositions 
of charged matter fields 
into Kaluza-Klein (KK) modes depend on the gauge
\cite{Hatanaka:1998yp}.
In this article, 
we elaborate this point in
a KK compactification
of massive Abelian gauge theory with charged matter fields.
After the KK dimensional reduction to 4D,
while whole KK mass spectrum is independent of the gauge,
the mode number depends on the gauge.
The masses of the KK modes of the charged matter fields
also depend on the field value of the zero-mode
of the Extra-Dimensional Component of the Gauge Field
(EDCGF for short in the following).
In particular, one mode in the KK tower 
in each massless 5D charged field
becomes massless at 
particular field values of the zero-mode of EDCGF.
Fields which become massless
at particular field values of inflaton 
lead to cosmological particle productions
which may have interesting observational consequences
\cite{Kofman:1997yn,Chung:1999ve,Kofman:2004yc,%
Barnaby:2009mc,Green:2009ds}.
In the meantime, the zero-mode of EDCGF
plays the role of inflaton in models of inflation
\cite{ArkaniHamed:2003wu,Kaplan:2003aj}.
In these models, 
the inflaton makes large field (trans-Planckian) excursion
to explain the observed anisotropy of the 
Cosmic Microwave Background (CMB) \cite{Ade:2015xua,Ade:2015lrj}.
If we consider our model in this context,
the inflaton passes through the
points at which the KK mode becomes massless
during its excursion, 
explaining the title of this article.
The mechanism explained in this article
is general in models of inflation
based on higher-dimensional gauge theories
as it is at work whenever
a light charged matter field is present in the model.

\section{KK modes of matter fields in 5D 
Abelian gauge theory}\label{sec:KK}

To present 
the basic mechanism
in the simplest example, 
we consider the following
5D action of massive gauge field
coupled to a massless scalar field:
\begin{align}
S_5 =
\int d^5 x&
\Bigl[
-\frac{1}{4} F_{MN}(x) F^{MN}(x)
- V(\cA_{M}(x)) + D_M {\chi}^{\dagger}(x) D^M \chi (x) 
\Bigr], 
\nn\\
&\qquad \qquad \qquad \qquad  (M,N = 0,1,2,3,5) ,
\label{action}
\end{align}
where
\begin{equation}
\cA_M(x) = A_M(x) - \frac{i}{g_5} e^{i \theta(x)} \pa_M e^{-i\theta(x)} ,
\label{cA}
\end{equation}
and $\theta (x)$ is the Stueckelberg field.
Note that the Stueckelberg field
is an angular variable
as seen from (\ref{cA}):
\begin{equation}
\theta(x) \sim \theta(x) + 2\pi w,
\label{ang}
\end{equation}
with $w$ being an integer and $\sim$ denotes the identification.
The field strength of the Abelian gauge field is given as
\begin{equation}
F_{MN} (x) = \pa_M A_N (x)  - \pa_N A_M (x) .
\label{FMN}
\end{equation}
We assume the minimal coupling 
of the field $\chi (x)$ to the gauge field:
\begin{equation}
D_M \chi (x) = \pa_M \chi (x)  - i g_5 A_M (x) \chi (x) .
\label{DMchi}
\end{equation}
The potential for the vector field
$V(\cA_M)$ may be expanded 
in power series in $\cA_M \cA^M$.
We take $V(\cA_M) = m^2 \cA_M \cA^M /2$ 
to make the following discussions simple,
but higher order terms can be easily included.
The action (\ref{action}) is invariant under the
local $U(1)$ gauge transformation
generated by
$U(x) = \exp [i \Theta(x)]$:
\begin{align}
A_M(x) \rightarrow&\,\,
\tilde{A}_M (x) = 
A_M(x) + \frac{i}{g_5} U(x) \pa_M U^{-1}(x),
\nn\\
\theta(x) \rightarrow&\,\, 
\tilde{\theta}(x) = 
\theta(x) + \Theta(x),
\nn\\
\chi(x) \rightarrow&\,\, 
\tilde{\chi}(x) = U(x) \chi(x),
\nn\\
\chi^\dagger (x)
\rightarrow&\,\,
\tilde{\chi}^\dagger(x) =
\chi^\dagger (x)
U^{-1}(x).
\label{gt}
\end{align}
Here, we indicated
by $\tilde{\ }$ that they are
the fields after the gauge transformation.

We are interested in the KK dimensional reduction
on a circle with radius $L_5$ in the 5-th direction.
We impose periodic boundary conditions on
the fields $\chi(x)$, $\chi^\dagger(x)$ 
and $A_M(x)$:\footnote{%
These fields can have boundary conditions
twisted by the global part of the gauge transformation,
but this case reduces to the periodic boundary conditions 
(\ref{pbc})
by field redefinitions.}
\begin{equation}
\psi (x^5 + 2\pi L_5) = \psi (x^5),
\quad
\psi(x) = A_M(x), \chi(x), \chi^\dagger(x).
\label{pbc}
\end{equation}
However,
the Stueckelberg field
can have a winding mode
as it is an angular variable as in (\ref{ang}):
\begin{equation}
\theta(x^5 + 2\pi L_5) 
= 
\theta(x^5) + {2\pi w}
\sim \theta(x),
\quad (w:\mbox{integer}).
\label{wind}
\end{equation}
To study KK dimensional reduction,
we decompose the field into Fourier modes:
\begin{align}
A_M (x,x^5)
&=
\frac{1}{\sqrt{2\pi L_5}}
\sum_{n=-\infty}^{\infty}
A_M^{(n)}(x)
e^{i \frac{n}{L_5} x^5},
\label{FA}\\
\theta(x,x^5)
&=
\frac{w x^5}{L_5}
+
\frac{1}{\sqrt{2\pi L_5}}
\sum_{n=-\infty}^{\infty}
\theta_n (x)
e^{i \frac{n}{L_5} x^5},
\quad
(w: \mbox{integer}),
\label{Fth}\\
\chi (x,x^5)
&=
\frac{1}{\sqrt{2\pi L_5}}
\sum_{n=-\infty}^{\infty}
\chi_n (x)
e^{i \frac{n}{L_5} x^5},
\label{Fchi}\\
\chi^\dagger (x,x^5)
&=
\frac{1}{\sqrt{2\pi L_5}}
\sum_{n=-\infty}^{\infty}
\chi_n^{\dagger} (x)
e^{-i \frac{n}{L_5} x^5} .
\label{Fchid}
\end{align}
From here we slightly change our notation:
$x^5$ is distinguished from 
$x$ which collectively denotes 4D Minkowski coordinates.
It is important to notice that the KK mode labels $n$ 
for the charged fields $\chi(x,x^5)$ and $\chi^{\dagger}(x,x^5)$
in (\ref{Fchi}) and (\ref{Fchid})
are gauge dependent, 
since they can be shifted 
by the gauge transformation 
(\ref{gt})
with the gauge parameter 
\begin{equation}
U(x,x^5) = e^{i \Theta(x,x^5)}, \quad \Theta(x,x^5) = \frac{x^5}{L_5}.
\label{Theta}
\end{equation}
Note that the gauge transformation
(\ref{Theta}) does not change the periodic boundary conditions
(\ref{pbc}) and (\ref{wind}) on the fields
and thus is a legitimate gauge transformation.
The gauge transformation (\ref{Theta}) shifts the Fourier modes
including the zero-modes of the fields as
\begin{align}
&\chi_n (x) 
\rightarrow 
\tilde{\chi}_n (x) = \chi_{n-1} (x),
\quad
\chi_n^\dagger (x) 
\rightarrow 
\tilde{\chi}_n^\dagger (x) = \chi_{n-1}^\dagger (x),
\label{modeshift}\\
&A_5^{(0)} (x) 
\rightarrow 
\tilde{A}_5^{(0)}(x) = A_5^{(0)} (x) + \frac{1}{g_4 L_5},
\quad
w \rightarrow \tilde{w} = w + 1.
\label{zeromodeshift}
\end{align}
In (\ref{zeromodeshift}) we have introduced
the four-dimensional gauge coupling $g_4$ 
which is 
related to the 5D one as
\begin{equation}
g_4 = \frac{g_5}{\sqrt{2\pi L_5}}.
\label{gg5}
\end{equation}
Some comments may be in order regarding (\ref{modeshift}).
From (\ref{gt}) the gauge transformation (\ref{Theta})
acts on the field $\chi(x,x^5)$ as follows:
\begin{equation}
\chi (x,x^5) \rightarrow \tilde{\chi}(x,x^5) 
= e^{i \frac{x^5}{L_5}} \chi(x,x^5) ,
\label{chigt}
\end{equation}
We write 
the Fourier decomposition of $\tilde{\chi}$ as follows:
\begin{equation}
\tilde{\chi} (x,x^5)
=
\frac{1}{\sqrt{2\pi L_5}}
\sum_{n=-\infty}^{\infty}
\tilde{\chi}_n (x)
e^{i \frac{n}{L_5} x^5},
\label{Ftchi}
\end{equation}
Then, through (\ref{chigt})
the Fourier modes $\chi_n (x)$ and $\tilde{\chi}_n (x)$ are related as
\begin{equation}
\tilde{\chi}_{n} (x) = \chi_{n-1} (x).
\label{chitchi}
\end{equation}

As shown below,
the 4D masses of the fields 
$\chi_n(x)$ and $\chi_n^\dagger(x)$
depend on the field value of $A_5^{(0)}$.
Unless we specify the field value of $A_5^{(0)}$,
we cannot tell
which mode is the lightest,
except which we wish to integrate out to obtain
the low energy effective theory below
the compactification energy scale $L_5^{-1}$.
Therefore, 
we keep all the Fourier modes of 
$\chi(x,x^5)$ and $\chi^\dagger(x,x^5)$
in KK dimensional reduction,
while keeping only the zero-modes
for other fields.\footnote{%
The necessity of this generalized 
KK dimensional reduction 
including all the KK modes of the charged fields
has been noted in \cite{Hatanaka:1998yp}.}
Upon KK dimensional reduction, 
we obtain the following 4D classical action:
\begin{align}
S_4
=
\int d^4 x
\Biggl[
&- \frac{1}{4} F_{\mu\nu} (x) F^{\mu\nu} (x)
+
\frac{m^2}{2}
\cA_\mu (x) \cA^\mu (x)
\Biggr.
\nn\\
&
+
\frac{1}{2} D_\mu \phi (x) D^\mu \phi (x)
- \frac{m^2}{2} 
\left( \phi (x) - 2 \pi f w \right)^2
\nn\\
\Biggl.
&+
\sum_{n=-\infty}^{\infty}
\left\{ 
D_\mu {\chi}_n^{\dagger} (x) D^\mu \chi_n (x)
+
\chi^\dagger_{n} (x)
g_4^2
\left(
\phi(x) - 2\pi f n
\right)^2
\chi_n (x)
\right\}
\Biggr],
\nn\\
& \qquad \qquad \qquad (\mu = 0,1,2,3),
\label{S4}
\end{align}
where we have defined 
\begin{equation}
2 \pi f = \frac{1}{g_4 L_5}.
\label{f}
\end{equation}
In (\ref{S4}),
we have omitted the suffix $0$
of the zero-modes, i.e. we have
renamed the fields as
$A_\mu^{(0)}(x) \Rightarrow A_\mu (x)$, 
$A_5^{(0)} (x) \Rightarrow \phi (x)$, 
$\theta_0 (x) \Rightarrow \theta (x)$,
and then we defined $\cA_\mu (x) = A_\mu (x) 
- i e^{i \theta(x)} \pa_\mu e^{-i \theta(x)}$.
The action is invariant under the
discrete shift (\ref{modeshift}), (\ref{zeromodeshift})
whose origin is the gauge symmetry (\ref{gt}):
\begin{align}
&\chi_n (x) \rightarrow \tilde{\chi}_n (x) = \chi_{n - 1} (x),
\quad
\chi_n^\dagger (x) \rightarrow 
\tilde{\chi}_n^\dagger (x) = \chi_{n-1}^\dagger (x),
\nn\\
&\phi (x)  \rightarrow \tilde{\phi} (x) = \phi (x) + 2 \pi f,
\quad
w \rightarrow \tilde{w} = w + 1 .
\label{shift2}
\end{align}
Note that after the KK dimensional reduction,
only the transformation of the winding mode
$w$ of the 5D field $\theta(x,x^5)$ remains.

Now, suppose that the field 
$\phi(x)$ at first had 
a vacuum expectation value 
$\langle \phi \rangle$,
then it makes
an excursion by $2\pi f$:
\begin{equation}
\langle \phi \rangle \longrightarrow \langle \phi \rangle + 2 \pi f.
\label{excursion}
\end{equation}
In the above we have used the long right arrow to denote
the field excursion.
We observe

1. The potential energy 
	 $V(\phi) = m^2 \left( \phi(x) - 2 \pi f w \right)^2/2$ 
	 changes as
	\begin{align}
  &V(\langle \phi \rangle) 
	= \frac{m^2}{2} \left( \langle \phi \rangle - 2 \pi f w \right)^2
	\nn\\
	\longrightarrow\,\, 
	&V(\langle \phi \rangle + 2\pi f)
	= \frac{m^2}{2} \left( \langle \phi \rangle - 2 \pi f (w-1) \right)^2.
	\label{Vf}
	\end{align}
	
2. The mass square of the fields $\chi_n(x)$ and $\chi_n^\dagger (x)$
	flows as
\begin{equation}
	g_4^2 \left(
\langle \phi \rangle - 2 \pi f n
\right)^2
  \longrightarrow
g_4^2 \left(
\langle \phi \rangle - 2\pi f (n-1)
\right)^2 ,
\label{massf}
\end{equation}

As noted in the first point, 
the value of the potential energy $V(\phi)$ changes
after the excursion (\ref{excursion}).
Using the inverse of the gauge transformation (\ref{shift2}),
one can identify the field value $\langle \phi \rangle + 2\pi f$ 
with the original value $\langle \phi \rangle$,
but then the winding number $w$ must be gauge transformed to $w-1$.
These mean that the system has undergone monodromy.
Models which realizes large field inflation using monodromy
have been extensively studied
started from \cite{Silverstein:2008sg,McAllister:2008hb}.
From the second point above, 
we observe that while 
the mass of each KK mode of the charged scalar field shifts 
by the excursion (\ref{excursion}),
the spectrum as a whole comes back to the original one.

Note that when the 4D gauge coupling $g_4 \lesssim 1$,
$2\pi f \gtrsim L_5^{-1}$ from (\ref{f}).
For applications to inflationary cosmology,
$L_5^{-1}$ should be larger than the Hubble scale
at the time of inflation.
Though such field excursion might look rather large,
large field excursions are very common
in inflation models,
for example in large field inflation models
inflaton makes trans-Planckian excursion.
We will present an example of
such models explicitly in section \ref{sec:obs}.

The interaction
between the field $\chi_n (x)$, $\chi_n^\dagger (x)$ 
and $\phi (x)$ 
in (\ref{S4})
is of the type
studied in 
\cite{Kofman:1997yn,Chung:1999ve,Kofman:2004yc,%
Barnaby:2009mc,Green:2009ds}.
In these models, 
at some value(s) of the inflaton field,
some field(s) become(s) massless and
cosmological particle productions follow,
which may have observational consequences.
In the current model,
when the field $\phi (x)$ passes through 
the field value $\phi = 2 \pi f n$
($n$: integer),
the modes $\chi_n (x)$ and $\chi_n^\dagger (x)$
become massless.
That
the field interval between the points
at which the matter fields become massless
are equally spaced
reflects the geometry of the compactified space
(circle), and it will be different for different compactified spaces.
The mass spectrum flow (\ref{massf}) follows
from the minimal coupling
and it is not restricted to scalar matter fields.
Similar mass spectrum flow
appears in models with fermionic matter fields massless in 5D.
As mentioned in the introduction,
applications of 
cosmological particle productions from
fields which become massless
at particular field values of inflaton
have been extensively studied following
\cite{Kofman:1997yn,Chung:1999ve,Kofman:2004yc,%
Barnaby:2009mc,Green:2009ds},
while models of inflation
in which the zero-mode of EDCGF plays the role of
inflaton have been extensively investigated
following
\cite{ArkaniHamed:2003wu,Kaplan:2003aj}.
Our study shows that such cosmological particle productions
are general feature of the inflation models
based on higher-dimensional gauge theories
which contain light charged fields,
and should be examined whenever one considers such models.

\section{Comparison of a simple model 
with observations}\label{sec:obs}

In this section we study 
a simple model which exhibits 
the resonant particle productions
discussed in the previous section
in light of recent CMB observations.
The purpose here is illustrative rather 
than comprehensive.
Some extensions of the model 
are discussed in the next section.

We consider a model in which
the zero-mode $\phi$ of EDCGF is identified with the inflaton.
We first assume that the back-reactions
of the resonant particle creations are small,
and later examine the 
parameter region where
this assumption is consistent.
Under this assumption,
the slow-roll parameters 
from the quadratic potential (\ref{Vf})
are calculated as
\begin{equation}
\epsilon(\phi)
:=
\frac{1}{2}\left(\frac{V'}{V}\right)^2
=
\frac{2}{\phi^2} ,\quad
\eta(\phi):=
\frac{V''}{V}
= \frac{2}{\phi^2}.
\label{epeta}
\end{equation}
Here, we work in the unit $M_P = 1$,
where $M_P := (8\pi G_N)^{-1/2}$
is the reduced Planck mass.
The spectral index is given as
\begin{equation}
n_s \simeq 1 - 6 \epsilon(\phi_\ast) + 2 \eta(\phi_\ast),
\end{equation}
where
the subscript $\ast$ refers to the value at
the pivot scale $0.002$ Mpc$^{-1}$ 
of the Planck 2015 analysis \cite{Ade:2015xua}.
The number of e-folds
is given as
\begin{equation}
N(\phi)
\simeq
\int_{\phi_{end}}^{\phi}
d\phi
\frac{V}{V'}
=
\left[
\frac{\phi^2}{4} 
\right]_{\phi_{end}}^{\phi},
\label{N}
\end{equation}
where we have defined $\phi_{end}$
by $\epsilon(\phi_{end}) = 1$.
This gives
\begin{equation}
\phi_{end} = \sqrt{2} .
\label{phiend}
\end{equation}
The scalar power spectrum is given by
\begin{align}
P_s
\simeq
\frac{V(\phi_\ast)}{24\pi^2 \epsilon(\phi_\ast)}
=
2.2 \times 10^{-9},
\label{Ps}
\end{align}
where the right hand side is 
the observed value \cite{Ade:2015xua}.
The tensor-to-scalar ratio is given as	
\begin{equation}
r_\ast \simeq 16 \epsilon (\phi_\ast) .
\label{r}
\end{equation}
Now we choose
$N_\ast =60$,
which 
gives 
\begin{align}
\phi_\ast &\simeq 16, \label{phi60}\\
r &\simeq 0.13, \label{r60}\\
n_s &\simeq 0.97, \label{ns60}\\
m &\simeq 6.0 \times 10^{-6}, \label{m60}\\
H_\ast &\simeq 5.8 \times 10^{-5}. \label{H60}
\end{align}
While the chaotic inflation with quadratic potential is
moderately disfavored
according to the Planck 2015 analysis \cite{Ade:2015lrj},
it is possible to 
improve the fit to the data
by a well-motivated modification of the potential
\cite{Furuuchi:2015jfj}.
Since our purpose here is to illustrate
the main idea in a simple model,
we proceed with the quadratic potential model.

There are also loop corrections
to the effective potential of the inflaton.
At one-loop,
the contributions from the charged matter fields
$\chi$ and $\chi^\dagger$
give \cite{Hatanaka:1998yp}
\begin{equation}
V_{1-loop}
=
\frac{3}{\pi^2(2\pi L_5)^4}
\left(
1 - \cos \frac{\phi}{f}
\right).
\label{Vloop}
\end{equation}
The effects of this term
on CMB anisotropy have been studied in
\cite{Flauger:2009ab,Flauger:2014ana}.
Here, we restrict ourselves 
to the parameter region where
these effects are small.
This is the case when
\begin{equation}
V' \gg
V_{1-loop}'
\simeq
\frac{3}{\pi^2 f (2\pi L_5)^4}.
\label{VcVl}
\end{equation}
Putting the field value 
$14 \lesssim \phi \lesssim 16$
during the observable inflation
and 
the inflaton mass (\ref{m60})
into (\ref{VcVl}),
we obtain
\begin{equation}
\frac{1}{L_5}
< g_4^{-1/3} \times 10^{-2}.
\label{Ccons}
\end{equation}
In order to justify the use of four-dimensional
Einstein gravity,
the Hubble expansion rate at the time of inflation
should be smaller than the compactification 
energy scale $1/L_5$.
From (\ref{H60}),
this condition gives
\begin{equation}
\frac{1}{L_5} > 6\times 10^{-5}.
\label{Hcons}
\end{equation}

In \cite{Barnaby:2009mc},
the peak amplitude 
in the CMB power spectrum 
from a single
isolated particle resonance 
due to the interaction between
the inflaton and
the charged fields $\chi$, $\chi^\dagger$
in the form in (\ref{S4})
was numerically obtained as
\begin{equation}
A_{IR} \simeq  2 \times g_4^{15/4} \times 10^{-6},
\label{AIR}
\end{equation}
(the factor $2$ in (\ref{AIR})
comes from the fact that $\chi$ and $\chi^\dagger$
are complex scalar fields).
We assume that the
contribution from
a single particle resonance
is smaller than the contributions of the
inflaton vacuum fluctuations (\ref{Ps}),
$P_s > A_{IR}$.
This condition gives
\begin{equation}
g_4 < 2 \times 10^{-1}.
\label{AIRcons}
\end{equation}

The number of e-folds between the
$i$-th particle resonance to $i+1$-th particle resonance
is given as
\begin{equation}
N(\phi(k_i)) - N(\phi(k_{i+1})
=
N(\phi(k_i)) - N(\phi(k_i) - 2\pi f)
=
\ln
\frac{k_{i+1}}{k_i} .
\label{DeltaN}
\end{equation}
During the observable inflation,
the Hubble expansion rate does not change much.
Therefore,
we can safely take the part linear in $2\pi f$
in the left hand side of (\ref{DeltaN}).
Then we obtain
\begin{equation}
k_{i+1} \simeq e^\Delta k_i,
\end{equation}
where
\begin{equation}
\Delta =
\frac{d N}{d\phi} 2\pi f
=
\frac{\phi}{2} 2 \pi f .
\label{Delta}
\end{equation}
Putting $\phi \simeq \phi_\ast \simeq 16$,
we obtain
\begin{equation}
\Delta \sim
\frac{8}{g_4 L_5}.
\label{Delta2}
\end{equation}
Then from
(\ref{Ccons}) and (\ref{Hcons})
we obtain
\begin{equation}
g_4^{-1} \times 5\times 10^{-4} 
< 
\Delta < g_4^{-4/3} \times 8 \times 10^{-2}.
\label{Delta3}
\end{equation}
For a reasonable value $g_4 = 0.1$, we have
\begin{equation}
5\times 10^{-3} 
< 
\Delta < 2.
\label{Delta4}
\end{equation}
(\ref{Delta4})
mostly falls in the case of
multiple bursts of particle production
in \cite{Barnaby:2009dd}.
Note that when $\Delta \ll 1$,
it becomes harder to observationally
distinguish each resonant particle production. 


\section{Note on models with 
more extra dimensions}\label{sec:moreX}

Instead of the 5D model discussed in 
the previous section,
we could have started with a gauge theory
with more extra dimensions.
In this case, 
we have multiple zero-modes 
of the
extra-dimensional components of the gauge field.
When the field space of the zero-mode of the gauge field
is one-dimensional like in the 
5D model studied in the previous section,
the field $\phi$ must pass through the point where
the KK modes of the charged fields
become massless
if it travels more than $2\pi f$.
On the other hand, 
if the field space of the zero-modes of the gauge field
has more than two dimensions,
the trajectory in the field space
need not pass through
the points where the KK modes of the 
charged fields 
become massless.
If one prefers to work in a model where
the KK modes of the charged fields
certainly become massless,
one may better assume that
the size of the 5-th dimension should be
hierarchically larger than the other extra dimensions
so that
below the compactification energy scale of other extra dimensions
the model effectively reduces to a 5D model.
Another choice for constructing
a model in which KK modes of charged field becomes light
may be to 
increase the number density 
of the points in the field space
where the KK modes of 
charged matter fields become massless,
so that the trajectory in the field space of
the zero-modes of the extra-dimensional components of the gauge field
is likely to pass close to these points.
Such models can be constructed as a simple extension
of the model discussed in this article.
For example, one can consider a product gauge group
$U(1)^r$ with sufficiently large
integer $r$ and arrange the model so that the
vacuum expectation values 
of the zero-modes of the
extra-dimensional components of the gauge field
distribute uniformly and dense in the field space of the zero-modes.
The zero-mode of EDCGF 
of one of the $U(1)$ or 
one linear combination of $U(1)$'s
may play the role of inflaton.
Then one can assign the $U(1)$ charges
so that when the inflaton field value coincides
with the expectation value of EDCGF of one $U(1)$
the KK mode of some charged matter field(s) become(s) massless.
One may further embed this model
into a model with larger gauge group $U(r)$
which is spontaneously broken to $U(1)^r$.
A system similar to this case has been studied in \cite{Green:2009ds}
but in a T-dual picture, 
which we discuss in the next section.

\section{T-dual picture}\label{sec:T-dual}

Some extension of the action (\ref{S4})
may arise as a low-energy effective field theory
of a D-brane system. 
That system would have T-dual description
in which 
the momentum numbers and the winding numbers
are interchanged
\cite{Kikkawa:1984cp}.
In the T-dual picture,
the role of the zero-mode of EDCGF
in the model discussed in this article
will be played instead by the 
distance between different D-branes,
and the Fourier mode number $n$ 
will be replaced by a winding number,
with the dual compactification radius 
$\tilde{L}_5 = \alpha'/L_5$.
Note however that when $L_5 \gg \alpha'^{1/2}$
the description based on the momentum modes
is simpler, as in the dual picture
the compactification radius is much smaller than the 
string scale and one has to deal with the full stringy description.
In other words,
two pictures are good descriptions 
in different regions of parameter space.
Also note that T-dual picture is available only when
the compactified space has a $U(1)$ isometry.
We can consider more general compactification 
of higher-dimensional gauge theories
on a manifold without $U(1)$ isometry.
Keeping these points in mind,
we observe that
the example studied in this article 
is similar to the examples studied in 
\cite{Kofman:2004yc,Green:2009ds}.

\section{Discussions}\label{sec:Discussions}

In this article 
we studied
KK dimensional reduction of higher-dimensional 
massive Abelian gauge theories
with charged matter fields,
and observed the appearance of 
massless fields
at particular field values of 
the zero-mode of EDCGF.
While in this article we took a simple model
as an example,
the appearance of massless fields at
particular field values of the zero-mode of 
EDCGF
is a general feature
in inflation models based on higher-dimensional gauge theories,
and should be examined further in such models.
It will particularly be interesting to study 
observational consequences
of the mechanism studied in this article
in explicit cosmology models.
It will also be interesting to study compactifications on
different manifolds.

\section*{Acknowledgments}
We would like to thank Yoji~Koyama
for the collaboration at the early stage of this work
as well as stimulating discussions and reading the manuscript. 

\bibliography{KKref}
\bibliographystyle{JHEP}
\end{document}